\documentclass[12pt]{article}
\usepackage{amssymb}

\newcommand{\comm}[2]{\left[#1,#2\right]}

\begin{document}
\title{Imaginary in all directions: an elegant formulation of special
relativity and classical electrodynamics}
\author{Martin Greiter and Dirk Schuricht\\[1cm]
Institut f\"ur Theorie der Kondensierten Materie,\\
  Universit\"at Karlsruhe, Postfach 6980, D-76128 Karlsruhe}
\maketitle
\pagestyle{plain}
\begin{abstract}
A suitable parameterization of space-time in terms of one complex and
three quaternionic imaginary units allows Lorentz transformations to
be implemented as multiplication by complex-quaternionic numbers
rather than matrices. Maxwell's equations reduce to a single
equation.
\end{abstract}

\section{Introduction}
One of the most established symmetries of our universe is the
invariance under Lorentz transformations, or the principle of special
relativity~\cite{wein}. Lorentz originally discovered that Maxwell's
theory of electromagnetism was invariant under these transformations.
Einstein subsequently interpreted them as a principle of relativity,
which is by no means specific to electromagnetism but a fundamental
and far reaching invariance of our universe. The transformations were
formulated in terms of $4\times4$ matrices
$(\Lambda^{\nu}_{\phantom{\nu}\mu})$ that act on contravariant four
vectors denoting space-time $(q^{\mu})\equiv (t,x,y,z)$,
energy-momentum $(p^{\mu})\equiv (E,p_x,p_y,p_z)$, or the like. The
introduction of the Minkowski metric $(g_{\mu\nu})$ and another set of
covariant vectors such that $(q_{\mu})\equiv (t,-x,-y,-z)$ provided a
highly convenient notation for relativistic theories, which has ever
since been taught and referred to as {\it relativistic notation}.
Einstein once even joked that his most important contribution to
physics was to introduce this notation.

Apart from the success of the notation, it is taught to students of
physics at such an early stage that they hardly question it. They
learn in other classes, for example in alternating current circuit
theory, the enormous technical advantage that can result if one
employs complex numbers to describe the phase shifts in the time
dependencies of real but alternating voltages and currents, but nothing
suggests that any of this formalism, or a generalization of it, can
simplify Lorentz transformations in a similar way. The only obvious
application of imaginary units appears to be a rotation from Minkowski
to Euclidean space by replacing the time $t$ by an imaginary time
$\tau\equiv -it$, with the effect that the Minkowski metric
$(g_{\mu\nu})$ is replaced by a simple Euclidean metric
$(\delta_{\mu\nu})$.

In this paper, we will show that a suitable imaginary parameterization
of space-time or other Lorentz contra- or covariant quantities, using
both one complex and three quaternionic imaginary
units~\cite{ham,dixon}, yields a tremendous simplification of
relativistic invariance~\cite{synge,leo,edm-book}: Lorentz
transformations will no longer require multiplication by matrices, but
only multiplication by complex-quaternionic (CQ) numbers.  Classical
electrodynamics will then provide us with a first orchard in which to
harvest the gain in formal elegance~\cite{Imaeda,edmonds}.  To be
specific, Maxwell's four equations will reduce to a single identity
between two CQ numbers.

\section{Special relativity}
Let us now set up the formalism.  We introduce a complex algebra with
generators $1,@ \in \mathbb{C}$, such that 
\begin{equation}
@^2=-1\;,
\end{equation} 
as well as an quaternionic algebra with generators 
$1, i, j, k \in\mathbb{H}$, such that 
\begin{equation}
\begin{array}{r@{\ }c@{\ }l}
&i^2=j^2=k^2=-1,&\\[1mm] 
&ij=-ji=k,\quad jk=-kj=i,\quad ki=-ik=j,&
\end{array}
\label{quaternions}
\end{equation}
which mutually commute:
\begin{equation}
\comm{i}{@}=\comm{j}{@}=\comm{k}{@}=0.
\end{equation}
We further introduce a complex conjugate operation $^*$, which takes
the form 
\begin{displaymath} 
{@} \rightarrow {@}^*=-{@} 
\end{displaymath} 
but leaves $i$, $j$, and $k$ unchanged, as well as a quaternionic
conjugate operation $^{-}$, which leaves @ unchanged but takes the form
\begin{displaymath}
i\rightarrow\bar i=-i,\quad j\rightarrow\bar j=-j,\quad
\hbox{and}\quad k\rightarrow \bar k=-k.  
\end{displaymath}  
Note that if $ o_1,o_2 \in\mathbb{C}\otimes\mathbb{H}\,$ are two CQ
numbers, the order of the product $o_1o_2$ is reversed under
quaternionic conjugation only:
\begin{equation}
(o_1o_2)^*=o_1^*o_2^*\qquad\hbox{but}\qquad \overline{o_1o_2}=\bar o_2\bar o_1.
\end{equation}
No simple product rule would exist if we had two or more mutually commuting 
quaternionic algebras.

We label space-time (and other Lorentz contravariant quantities usually 
denoted by four vectors) by a purely imaginary CQ number, 
\begin{equation}
q\equiv @t+ix+jy+kz
\label{eq:im}
\end{equation}
where $t,x,y,z \in \mathbb{R}$. We identify this subspace of
$\mathbb{C}\otimes\mathbb{H}$ using Minkowski space and denote it by
$\mathbb{M}$. Complex conjugation $^*$ and quaternionic conjugation
$^{-} $ correspond within this space to time reversal (T) and parity
(P) transformations, respectively. Note that $q^*=-\bar q$ for
$q\in\mathbb{M}\;\!$.

The corresponding covariant quantity is given by its quaternionic 
conjugate or parity reversed CQ number,
\begin{equation}
\bar q= @t-ix-jy-kz,
\end{equation}
yielding the proper time interval
\begin{equation}
-\bar q q = - q \bar q = t^2-x^2-y^2-z^2.
\label{eq:propertime} 
\end{equation}

Let $n=in_x+jn_y+kn_z$ with $n \bar n = n_x^2+n_y^2+n_z^2=1$ and 
$ n_x,n_y,n_z \in\mathbb{R}$ be a quaternionic imaginary unit
vector $(n_x,n_y,n_z)$. Then a Lorentz transformation is given simply by
\begin{equation}
q\rightarrow q'=\omega q\bar\omega^*,
\label{eq:lt}
\end{equation}
with either
\begin{equation}
\omega= e^{\frac{1}{2}n\theta} = 
\cos\frac{\theta}{2}+n\sin\frac{\theta}{2}
\label{eq:rot}
\end{equation}
for a rotation by an angle $\theta$ around $n$, or 
\begin{equation}
\omega= e^{\frac{1}{2} n @ \Lambda} = 
\cosh\frac{\Lambda}{2}+@n\sinh\frac{\Lambda}{2}
\label{eq:boo}
\end{equation}
for a boost by a Lorentz angle $\Lambda$ in direction $n$.
(\ref{eq:lt}-\ref{eq:boo}) are readily verified using $n^2=-1$ and
$(n@)^2=1$.

Clearly the covariant CQ number $\bar q$ transforms as 
\begin{equation}
\bar q\rightarrow \bar q'=\omega^* \bar q\bar\omega.
\end{equation}
With $\omega \bar \omega = \bar\omega\omega=1$, 
the Lorentz invariance of the proper time $-\bar qq$ is evident.

At this point we assume that the reader is aware of how awkward it is
to write down a rotation around or a boost along an arbitrary axis using
$4\times4$ matrices, so that there is no need to dwell on the elegance
of the formulation proposed here.

It is convenient at this point to use the norm given by the proper
time interval to define a scalar product between two imaginary or 
Minkowski CQ numbers $\mathbb{M} \times
\mathbb{M} \rightarrow\mathbb{R}$, according to 
\begin{eqnarray} 
\langle p,q\rangle &\equiv &\frac{1}{4}
\left(\overline{(p-q)}(p-q)-\overline{(p+q)}(p+q)\right)\nonumber\\
&=&-\frac{1}{2}(\bar pq+\bar qp)\nonumber\\
&=&-\frac{1}{2}(p\bar q+q\bar p).
\label{eq:scp} 
\end{eqnarray} 
The proper time interval (\ref{eq:propertime}) is 
given by $\langle q,q\rangle =-\bar q q$.
With $p=@E+ip_x+jp_y+kp_z$ and $q=@t+ix+jy+kz$, we obtain
\begin{equation} 
\langle p,q\rangle\, = Et-p_xx-p_yy-p_zz .
\label{eq:pdotq} 
\end{equation}

Furthermore, it is convenient to define the contravariant differentiation
operator
\begin{equation} 
D \equiv @ \frac{\partial}{\partial t}
-i\frac{\partial}{\partial x} - j\frac{\partial}{\partial y} -
k\frac{\partial}{\partial z} 
= @\partial_t -i\partial_x -j\partial_y -k\partial_z,
\label{eq:D} 
\end{equation}
which likewise transforms according to
\begin{equation}
D\rightarrow D'=\omega D\bar\omega^*.
\end{equation}
Note that $-D \bar q=-\bar D q=4$, as expected. The covariant
differentiation operator is, of course, given by
\begin{equation} 
\bar D = @\partial_t +i\partial_x +j\partial_y +k\partial_z,
\label{eq:Dbar} 
\end{equation}
and transforms  
\begin{equation}
\bar D\rightarrow \bar D'=\omega^* \bar D\bar\omega.
\end{equation}

\section{Classical electrodynamics}
We proceed by applying this formalism to classical electrodynamics.
To begin with, we introduce a contravariant vector (or imaginary CQ 
field)
\begin{equation}
A\equiv @\phi+iA_x+jA_y+kA_z,
\label{eq:A}
\end{equation}
and require the theory to be invariant under electromagnetic gauge
transformations 
\begin{equation}
A\rightarrow A + D\lambda(q),
\label{eq:gaugetrans}
\end{equation}
where $\lambda(q)$ is an arbitrary real-valued scalar function of
space-time $q$. We proceed by defining the electromagnetic field
strength
\begin{equation}
F\equiv \frac{1}{2}\left(\bar DA-\overline{\bar DA}\right).
\label{eq:F}
\end{equation}
Clearly $F$ is invariant under (\ref{eq:gaugetrans}). Note that we cannot
replace $\overline{\bar DA}$ by $\bar AD$ in the second term in (\ref{eq:F}),
since the derivative operator has to act on $A$. Under a Lorentz 
transformation, $F=-\bar F$ transforms as 
\begin{equation}
F\rightarrow F'=\omega^*F\bar\omega^*.
\end{equation}
At this point it is propitious to introduce
\begin{equation}
e_1\equiv i,\quad e_2\equiv j,\quad e_3\equiv k
\label{eq:eidef}
\end{equation}
such that 
\begin{equation}
e_i e_j = -\delta_{ij} + \epsilon_{ijk} e_k
\quad\hbox{with}\quad i,j,k\in\{1,2,3\},
\label{eq:eiej}
\end{equation}
where the indices $i,j,k$ in (\ref{eq:eiej}) are not to be confused with the 
quaternionic generators $i,j,k$ in (\ref{eq:eidef}). With summation over
repeated indices but no hidden minus signs implied, we write 
\begin{equation}
A= @\phi+e_iA_i\qquad D=@\partial_t-e_i\partial_i\qquad \hbox{etc.}
\label{eq:rewrite}
\end{equation}
Writing the field strength (\ref{eq:F}) out in components, we obtain
\begin{equation}
F=e_i \epsilon_{ijk} \partial_j A_k - @e_i (-\partial_t A_i-\partial_i \phi). 
\label{eq:Fexpl}
\end{equation}
Note that 
\begin{equation}
DF=(-@\partial_i+e_i\partial_t)(-\partial_t A_i-\partial_i \phi)-
e_i\epsilon_{ijk}\partial_j \epsilon_{klm}\partial_l A_m
\label{eq:DF}
\end{equation}
is a purely imaginary CQ number ($DF\in\mathbb{M}$). This implies  
\begin{equation}
DF+\overline{DF}^*=0,
\label{eq:max1}
\end{equation}
which will prove useful below.

We proceed by defining magnetic and electric field strengths according
to
\begin{equation}
B_i\equiv \epsilon_{ijk} \partial_j A_k \quad\hbox{and}\quad 
E_i\equiv -\partial_t A_i-\partial_i \phi,
\label{eq:BE}
\end{equation}
and use them to rewrite (\ref{eq:Fexpl}) as
\begin{equation}
F=e_i (B_i-@ E_i).
\label{eq:FBE}
\end{equation}
We may hence write the Lagrangian density for the electromagnetic
field coupled to an external current $(\rho,J_i)$ as
\begin{eqnarray}
\mathcal{L}&\equiv &\frac{1}{2}\left(E_i^2-B_i^2\right)+\rho\phi-J_iA_i
\nonumber\\
&=&\frac{1}{4}\left(F^2+(F^*)^2\right)+
\frac{1}{2}\left(\bar JA+\bar AJ\right),
\label{eq:L}
\end{eqnarray}
where we have defined
\begin{equation}
J\equiv @\rho + e_i J_i.
\label{eq:J}
\end{equation}
Clearly, $\mathcal{L}$ and $F^2=E_i^2+2@E_iB_i+B_i^2$ are Lorentz invariant.

Variation of $\mathcal{L}$ with respect to $\delta\bar A$ yields,
after integration by parts,
\begin{equation}
\frac{1}{4}\big(DF-\overline{DF}^*\big)+\frac{1}{2}J=0.
\end{equation}
Together with (\ref{eq:max1}), we obtain
\begin{equation}
DF+J=0.
\label{eq:max}
\end{equation}
This is {\it Maxwell's equation}. (In the standard formulation, there
is need to speak of {\it equations} (plural), but we shall see now
that this single equation replaces all four of them.)  Note that
(\ref{eq:max}) transforms contravariantly under Lorentz
transformations. Evaluation of $DF$, starting from (\ref{eq:FBE}),
yields
\begin{equation}
DF=\partial_iB_i-@\partial_iE_i+
e_i\left(\partial_tE_i-\epsilon_{ijk} \partial_j B_k\right)+
@e_i\left(\partial_tB_i+\epsilon_{ijk} \partial_j E_k\right).
\label{eq:DFEB}
\end{equation}
Since (\ref{eq:max}) implies that the coefficients of $DF+J$ in 
all eight orthogonal directions $\{1,@,e_i,@e_i\}$ of 
$\mathbb{C}\otimes\mathbb{H}$ vanish, substitution of (\ref{eq:DFEB})
and (\ref{eq:J}) into (\ref{eq:max}) immediately yields
\begin{equation}
\begin{array}{r@{\ }c@{\ }l}
\partial_iB_i&=&0\\[1mm]
-\partial_iE_i+\rho&=&0\\[1mm]
\partial_tE_i-\epsilon_{ijk} \partial_j B_k+J_i&=&0\\[1mm]
\partial_tB_i+\epsilon_{ijk} \partial_j E_k&=&0.
\end{array}
\label{eq:maxold}
\end{equation}
We assume the reader is familiar with these equations.

In Lorentz gauge, 
\begin{equation}
\partial_t \phi - \partial_i A_i =-\frac{1}{2} 
\left(\bar D A + \overline{\bar D A}\right) = 0.
\label{eq:lorentzgauge}
\end{equation}
Maxwell's equation (\ref{eq:max}) then reduces to 
\begin{equation}
D\bar D A +J=0, 
\label{eq:maxpotential}
\end{equation}
which is readily recognized as a wave equation.

\section{Conclusion}
In conclusion, we have shown that a suitable CQ 
parameterization of Lorentz contra- and covariant quantities 
can greatly enhance elegance and simplicity of relativistic theories.
We believe that this language is likely to yield new perspectives on
quantum field theories~\cite{morita,sharma,adler,moreleo}, 
which we are currently investigating.

\section*{Acknowledgment}
One of us (MG) is deeply indebted to S C Zhang~\cite{zhang} for 
countless discussions
on his four-dimensional generalization of fractionally quantized Hall
states that have inspired the present work.

\end{document}